\documentclass[aps, a4paper, prb, twocolumn]{revtex4-2}

\usepackage{amsmath}
\usepackage{graphicx}
\usepackage{bm}
\usepackage{xcolor}
\usepackage[caption=false]{subfig}

\newcommand{\ep}{\epsilon}
\newcommand{\bpm}{\begin{pmatrix}}
\newcommand{\epm}{\end{pmatrix}}

\newcommand{\fe}{\mathcal{F}}

\begin{document}

\title{Shear Modulus Anomaly of Unconventional Superconductor in a Symmetry Breaking Field}

\date{\today}
\author{Pye Ton How}
\affiliation{Physics Division, National Center for Theoretical Sciences, Hsinchu 300, Taiwan}

\author{Sung-Kit Yip}
\affiliation{Institute of Physics, Academia Sinica, Taipei 115,
	Taiwan}
\affiliation{Institute of Atomic and Molecular Sciences, Academia
	Sinica, Taipei 106, Taiwan}
\affiliation{Physics Division, National Center for Theoretical Sciences, Hsinchu 300, Taiwan}

\begin{abstract}
Using Ginzburg-Landau formalism, we theoretically study the isothermal shear modulus anomaly of an unconventional superconductor with a multicomponent order parameter, when the superconducting transition is split by a symmetry-breaking field.  Experimental signatures are proposed for both chiral and nematic superconductors.  Particularly striking is the vanishing of $C_{66}$ across the lower transition to a nematic superconducting state.  Our findings can guide future experiments and shed new lights on materials such as Sr$_2$RuO$_4$ and $M_x$Bi$_2$Se$_3$.
\end{abstract}

\maketitle

A superconductor described by a multi-component order parameter falls under the category of unconventional superconductivity\cite{Sigrist1991}.  In the presence of a symmetry-breaking field (SBF) that explicitly breaks the lattice symmetry, the superconducting transition is split into two branches\cite{Sigrist1987a, Volovik1988}.  UPt$_3$ is the earliest material that has lead to the conception of this scenario, where the SBF is an SDW order above superconductivity\cite{Fisher1989, Machida1989, Hess1989, Blount1990, Agterberg1995, Joynt2002}.

Controversial abounds, however, around Sr$_2$RuO$_4$ and $M_x$Bi$_2$Se$_3$, two other materials that may fit into this scenario.  The superconductivity of Sr$_2$RuO$_4$ appears to break the time-reversal symmetry (TRS)\cite{Luke1998}.  Chiral superconductivity seems a natural interpretation\cite{Sigrist1999}, but the associated edge current has been consistently not observed\cite{Kirtley2007, Hicks2010, Kashiwaya2011, Curran2014}, casting doubt on the scenario.  Experiments with uniaxial strain applied as an artificial SBF turn out mixed results: no thermodynamic signature for the split transitions is observed\cite{Watson2018,Li2019}, yet in a $\mu$-SR experiment\cite{Grinenko2020, Romer2020} the breaking of TRS occurs below the superconducting $T_c$, supporting the prediction of split transitions.

The nematic superconductivity model of $M_x$Bi$_2$Se$_3$\cite{Fu2014} necessitates a yet-unidentified pre-existing ``pinning field''\cite{How2019} to explain the persistent nematic orientation of any given sample in repeated experiments\cite{Kuntsevich2018, Kuntsevich2019}.  A very small distortion of crystal lattice has been observed\cite{Kuntsevich2018, Kuntsevich2019, Frohlich2020}, and it is suggested that this distortion accounts for the pinning effect\cite{Kuntsevich2018, Kuntsevich2019}.  Current body of experiemtnal evidence supporting this so-called ``nematic hypothesis'' consists of large two-fold anisotropic responses incompatible with lattice symmetry\cite{Matano2016, Pan2016, Asaba2017, Du2017, Yonezawa2017, Smylie2018, Willa2018, Sun2019, Fang2020} that emerge with the superconductivity.  But this is not logically inevitable: the pinning field alone provides the two-fold direction that breaks the trigonal lattice symmetry, regardless of the nature of the order parameter itself.  The pair of split transitions remains the only definitive evidence for multi-component superconductivity.  To our knowledge, none of the hallmarks\cite{Sigrist1987a, Volovik1988,How2019} pertaining to the split transitions have been tested for in existing experiments.

Recent ultrasound experiments reveal shear modulus anomaly in Sr$_2$RuO$_4$\cite{Benhabib2020,Ghosh2020} across the superconducting transition in the absence of an SBF, supporting the hypothesis of a multi-component order parameter. We hope that ultrasound measurement of the shear moduli will be a fruitful direction in the search of split transitions.  Taking the Ginzburg-Landau (GL) theory of a two-component order parameter as our starting point, we examine the discontinuity in isothermal shear modulus across the two branches of transitions.  We find that, in particular, one of the shear moduli \emph{vanishes} at the lower transition of $M_x$Bi$_2$Se$_3$.  Such a drastic signal should be readily observed in an experiment.  Prior experimental work on UPt$_3$ exists\cite{Thalmeier1991} but, as we will presently explain, we disagree with their theoretical analysis.

We will start by introducing the GL theory of a two-component order parameter, and coupling it to the planer shear strain and an SBF.  Next, we evaluate the shear moduli, paying particular attention to their qualitative behavior across each of the split transitions.  There are three separate cases: upper transition, lower transition into the chiral phase, lower transition into another nematic phase.  We then proceed to introduce the modified GL theory that describe nematic superconductivity in a trigonal crystal (e.g. $M_x$Bi$_2$Se$_3$) and work out the behavior of the shear moduli for the case.  Finally we discuss the implication of our result on the verification of split superconducting transition in such materials.


\emph{Ginzburg-Landau theory.}  For the three materials highlighted in the introduction (UPt$_3$, Sr$_2$RuO$_4$ and $M_x$Bi$_2$Se$_3$), the lattice point groups are respectively $D_{6h}$, $D_{4h}$ and $D_{3d}$.  They all admit two-dimensional irreducible representations.  We assume the (complex) superconducting order parameter $\vec{\eta}=(\eta_x, \eta_y)$ transforms in the said representation in each case.  The GL free energy for this order parameter is conventionally written as
\begin{equation}
\mathcal{F}_0 = a\vert\eta\vert^2 + b_1 \vert\eta\vert^4
	+ \frac{b_2}{2}\left[(\eta_x^*\eta_y)^2 + (\eta_y^*\eta_x)^2 \right]
	+ b_3 \vert\eta_x\vert^2\vert\eta_y\vert^2.
	\label{GLoriginal}
\end{equation}
This can be cast in a form that more cleanly displays the physics under discussion.  Let $\lbrace\sigma_i, i = 1,2,3 \rbrace$ be the usual Pauli matrices, and define $\langle\sigma\rangle_i \equiv (\vec{\eta}^{\dag} \sigma_{i-1} \vec{\eta})$ (identifying $\sigma_{1-1} = \sigma_3$):
\begin{equation}
\begin{split}
\fe_0 &= a\vert\eta\vert^2
	+ \frac12 \sum_{i=1}^{3} \Lambda_i \langle\sigma\rangle_i^2;\\
\Lambda_1 &= 2b_1, \quad \Lambda_2 = 2b_1 + \frac12 (b_3 + b_2),\\
\Lambda_3 &= 2b_1 + \frac12 (b_3 - b_2).	
\end{split}
\label{GLsigma}
\end{equation}
Throughout this paper, there will be \emph{no summation over repeated indices} unless explicitly indicated.

The theory \eqref{GLsigma} has exactly one critical point at $a = 0$.  Below the critical point, the smallest $\Lambda_i$ among the three decides the equilibrium sperconducting state: $\vec{\eta}$ is an eigenvector of $\sigma_{i-1}$.  The nematic state with a real $\vec{\eta}$ requires $i= 1$ or $2$, while $i = 3$ yields the complex chiral state.

We introduce a background SBF $\Delta$ that favors a real $\vec{\eta}$ through a coupling term in free energy:
\begin{equation}
\fe_{SBF} = -\Delta \langle\sigma\rangle_1.
\label{GLpinning}
\end{equation}
This was first proposed to explain the split transitions of UPt$_3$\cite{Hess1989, Machida1989, Blount1990}.  For $M_x$Bi$_2$Se$_3$, the reported pinning field is along either the $a$ or $a^*$ lattice direction for each individual sample\cite{Kuntsevich2019, Kawai2020}, and \eqref{GLpinning} is also appropriate.  For Sr$_2$RuO$_4$, this SBF may be applied artificially, for instance using uniaxial strain technique.  Throughout this paper we assume $\Delta > 0$.  The opposite case can be trivially obtained with some sign changes.

The SBF explicitly breaks the crystal symmetry down to $D_{2h}$.  The (upper) critical temperature is raised to $a_{u} = \vert\Delta\vert > 0$, and the order parameter is \emph{pinned} along $(1,0)$, regardless of the preference of free energy \eqref{GLsigma}.

Should $\Lambda_1$ not be the smallest, the originally preferred state eventually becomes competitive at a lower temperature, and the \emph{lower transition} occurs.  If $\Lambda_2$ is smallest, $\vec{\eta}$ remains real but tilts away from $(1,0)$, signaling the breaking of the horizontal two-fold rotational symmetry.  If $\Lambda_3$ is smallest, $\vec{\eta}$ becomes a combination of $(1,0)$ and an isotropic chiral component $(1,\pm i)$, breaking the TRS but preserving the $D_{2h}$ (the horizontal two-fold rotation must be redefined to be followed by time-reversal).  If $\Lambda_i$ ($i \neq 1$) is the smallest, the lower critical temperature is
\begin{equation}
a_l = -\left(\frac{\Lambda_i}{\Lambda_1-\Lambda_i} \right)\Delta < 0.
\label{TcLower}
\end{equation}

Up to an overall phase, $\vec{\eta}$ is parameterized as
\begin{equation}
\vec{\eta} = \vert\eta\vert \bpm \cos \theta \\
						e^{i\chi} \sin \theta \epm.
\end{equation}
The equilibrium solution can be found.  In the upper phase $a_u > a > a_l$:
\begin{equation}
\vert\bar{\eta}\vert^2 = \frac{\Delta-a}{\Lambda_1}, \quad \bar{\theta} = 0, \quad \chi\; \text{drops out.}
\label{upperEquilibrium}
\end{equation}
The specific heat discontinuity across the upper transition is
\begin{equation}
C_{\text{upper}} - C_{\text{normal}} = (T_0+\Delta)/\Lambda_1,
\label{specificHeatJumpUpper}
\end{equation}
where $T_0$ is the critical temperature of the single transition in the absence of any SBF.  It is related to $a$ by $a = (T-T_0)$.

For $a < a_l$, the lower phase solution is:
\begin{equation}
\begin{split}
\vert\bar{\eta}\vert^2 &= -\frac{a}{\Lambda_i}, \quad
\vert\bar{\eta}\vert^2 \cos 2\bar{\theta} = \left(\frac{\Delta}{\Lambda_1-\Lambda_i}\right), \quad \\
\bar{\chi} &= \begin{cases}
0 &\quad (i = 3, \text{nematic})\\
\pm\pi/2 &\quad (i = 2, \text{chiral}).
\end{cases}
\end{split}
\label{lowerEquilibrium}
\end{equation}
For the lower transition into either phase, $\theta$ serves as an order parameter.  It exhibits the usual meanfield critical behavior $\bar{\theta} \propto \pm \vert a - a_l\vert^{1/2}$ for $a < a_l$.  The specific heat jump across the lower transition is:
\begin{equation}
C_{\text{lower}} - C_{\text{upper}} = (T_0 + a_l)\left(\frac{1}{\Lambda_i}-\frac{1}{\Lambda_1} \right).
\label{specificHeatJumpLower}
\end{equation}

We next consider the coupling exclusively to shear strain in the basal plane:
\begin{equation}
\ep_{1} = \ep_{xx} - \ep_{yy}, \quad \ep_{2} = 2\ep_{xy}.
\end{equation}
These forms a 2D representation under $D_{3d}$ or $D_{6h}$. Under $D_{4h}$, they are separate one-dimensional representations.  These are introduced to the GL theory by adding
\begin{equation}
\fe_\ep = -g_1 \ep_1 \langle\sigma\rangle_1
	- g_2 \ep_2 \langle\sigma\rangle_2
	+ \frac{c_1}{2} \ep_1^2 + \frac{c_2}{2} \ep_2^2.
	\label{GLstrain}
\end{equation}
Here $c_1$ and $c_2$ are shear moduli in the normal phase.

The coupling to the strain renormalizes the quartic coefficients\cite{Benhabib2020} when the strain is eliminated using GL equations.  It is therefore necessary to replace the quartic coefficients in \eqref{GLoriginal} with the ``bare'' ones: define
\begin{equation}
\lambda_3 = \Lambda_3; \qquad\lambda_i = \Lambda_i + \frac{g_i^2}{c_i}, \quad i = 1, 2. 
\label{renormalization}
\end{equation}
We expect this correction to be quite small\footnote{The jumps of elastic moduli across the unsplit (or upper) transition indicates the size of the renormalization effect \eqref{renormalization} (c.f. equation \eqref{upperStep}, and also Ref \cite{Benhabib2020, Ghosh2020, Thalmeier1991}), and is usually found in the range of $10^{-4}$\cite{Benhabib2020, Ghosh2020, Thalmeier1991}.  In addition, the Testardi thermodynamic relation\cite{Testardi1971} misses this renormalization\cite{Note3}, but still works well for most cases.  For instance, in the supplemental material of ref \cite{Benhabib2020}, the relation is tested out on Sr$_2$RuO$_4$ and produces good agreement.}.  But it will prove to be qualitatively important in the case of trigonal crystal.

At the end, the free energy under consideration is
\begin{equation}
\begin{split}
\fe &= a\vert\eta\vert^2 - \Delta \langle\sigma\rangle_1
		+\frac{1}{2}\sum_{i = 1}^{3} \lambda_i\langle\sigma\rangle_i^2 \\
& \qquad+ \sum_{i=1,2} \left(-g_i \ep_i \langle\sigma\rangle_i + \frac{1}{2} c_i \ep_i^2 \right).
\end{split}
\label{GLfull}
\end{equation}
The equilibrium strain that satisfies the condition $\partial \fe/\partial \ep_i = 0$ is
\begin{equation}
\bar{\ep}_i = \frac{g_i}{c_i} \langle \sigma\rangle_i, \quad i = 1, 2.
\end{equation}
Once this condition is imposed, the theory \eqref{GLfull} retains the same equilibrium solutions \eqref{upperEquilibrium} and \eqref{lowerEquilibrium}.

For the trigonal and hexagonal cases ($D_{3d}$ and $D_{6h}$), the symmetry forces $\Lambda_1 = \Lambda_2$, $g_1 = g_2$ and $c_1 = c_2$.  The free energy \eqref{GLoriginal} becomes invariant under arbitrary rotation about the principal axis.  To describe the lower transition into nematic phase for these crystals, the emergent $O(2)$ symmetry needs to be broken down to a hexagonal one by the addition of sixth order terms in the free energy; this will be discussed later.


\emph{Upper transition.}  Let us first analyze the normal-to-superconducting transition at $a = \Delta$, specifically the effect on the elastic moduli.  To be precise, let $F(\ep_1, \ep_2) = \fe$ with the condition $\partial \fe/ \partial \vec{\eta} = 0$ enforced, and define $c_{i,j} = \partial^2 F/\partial\ep_i\partial\ep_j$, to be evaluated at equilibrium\footnote{See supplemental material for our own approach}.  The modulus $c_{i,j}$ can be expressed in Voigt notation as:
\begin{equation}
    \begin{split}
        c_{1,1} &= \frac{1}{4}\left(C_{11}+C_{22}-2C_{12}\right), \\
        c_{2,2} &= C_{66},\\
        c_{1,2} &= \frac{1}{2}\left(C_{16} - C_{26}\right).
    \end{split}
\end{equation}
It is obvious that in the normal phase $c_{1,1} = c_1$, $c_{2,2} = c_2$ and $c_{1,2} = 0$.

One needs to solve $\partial\fe/\partial\vec{\eta}$ for arbitrary $\ep_1$ and $\ep_2$ near equilibrium.  Given the equilibrium $\bar{\ep}_2 = 0$ in the upper phase, a small-$\ep_2$ expansion to leading order suffices.  The perturbed solution is
\begin{equation}
\vert\tilde{\eta}\vert^2 \approx \frac{g_1 \ep_1 + \Delta - a}{\lambda_1}; \quad
\tilde{\theta} \approx \frac{g_2 \ep_2/2}{g_1 \ep_1 + \Delta - (\lambda_1 - \lambda_2)\vert\bar{\eta}\vert^2},
\end{equation}
and $\chi = 0$.  (While $\bar{\eta}^2$ does admit correction at $O(\ep_2^2)$, the very condition $\partial \fe/\partial \vert\eta\vert = 0$ ensures that $c_{22}$ is independent of this correction term.)  The upper phase moduli are found to be:
\begin{equation}
\begin{split}
c^{(u)}_{1,1} &= c_1 - \frac{g_1^2}{\lambda_1};\\
c^{(u)}_{2,2} &= c_2 - \frac{g_2^2 (\Delta-a)}{(\lambda_2 - \Lambda_1)(\Delta-a) + \Lambda_1 \Delta },
\end{split}
\label{upperModuli}
\end{equation}
and the off-diagonal $c^{(u)}_{1,2} = 0$ as required by the $D_{2h}$ symmetry.

The modulus $c_{1,1}$ is discontinuous across the upper transition:
\begin{equation}
\delta_u c_{1,1} \equiv c^{(u)}_{1,1} - c_1 = - \frac{g_1^2}{\lambda_1},
\label{upperStep}
\end{equation}
but $c_{2,2}$ remains continuous.  However, in the limit of $\Delta \rightarrow 0$, $c_{2,2}$ develops a step discontinuity, too: the presence of $\Lambda_1\Delta$ in the denominator smooths out the would-be singular behavior.

\begin{figure}
\includegraphics[width = 0.48\textwidth]{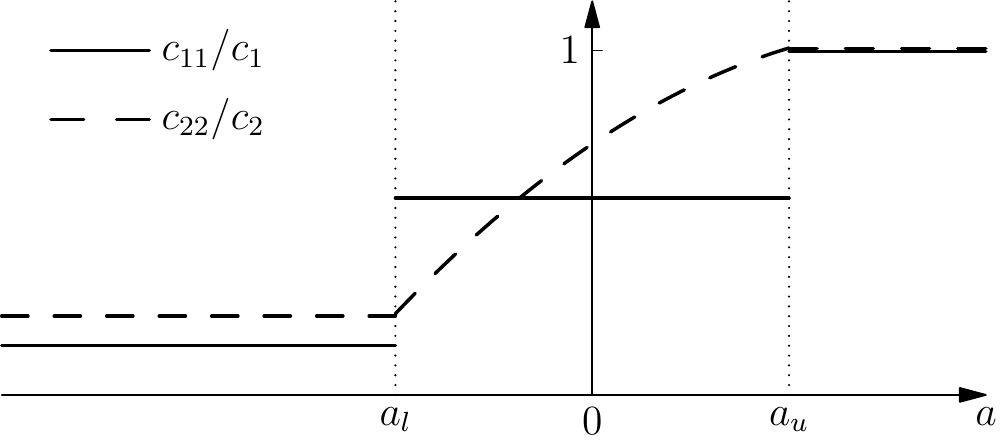}
\caption{\label{moduliPlotChiral} Qualitative behaviors of the dimensionless $c_{i,i}/c_i$ across the pair of split transitions, assuming the lower phase is chiral.  If the crystal is trigonal or hexagonal (e.g. UPt$_3$), one has $c_{1,1}/c_1 = c_{2,2}/c_2$ in the lower phase.  Otherwise the relative heights of the low-temperature plateaus are dependent on GL parameters.}
\end{figure}

\emph{Lower transition into chiral phase.}  Assuming that $\Lambda_3$ is the smallest, the order parameter acquires a chiral component when $a < a_l$.  The equilibrium solution is given in \eqref{lowerEquilibrium}.  We carry out the same procedure to extract the elastic moduli in the lower phase.  Given $\bar{\ep}_2 = 0$, we again expand only to linear order in $\ep_2$ to obtain (picking the positive branch for $\chi$):
\begin{equation}
\begin{split}
\vert\tilde{\eta}\vert^2 &= -\frac{a}{\Lambda_3}, \quad
\vert\tilde{\eta}\vert^2 \cos 2\tilde{\theta} = \left(\frac{g_1\ep_1+\Delta}{\lambda_1-\Lambda_3}\right), \\
\tilde{\chi} &= \frac{\pi}{2} + \frac{g_2 \ep_2}{(\Lambda_3 -\lambda_2)\vert\tilde{\eta}\vert^2\sin 2\tilde{\theta}}.
\end{split}
\end{equation}
The lower chiral phase moduli are
\begin{equation}
c^{(c)}_{1,1} = c_1 - \frac{g_1^2}{\lambda_1 - \Lambda_3}, \quad
c^{(c)}_{2,2} = c_2 - \frac{g_2^2}{\lambda_2 - \Lambda_3},
\label{chiralModuli}
\end{equation}
and $c^{(c)}_{1,2} = 0$ again, as dictated by the $D_{2h}$ symmetry.  Compared with the upper phase result \eqref{upperModuli}, at the lower transition $c_{1,1}$ is discontinuous and $c_{2,2}$ is continuous with a kink:
\begin{equation}
\delta_{c} c_{1,1} \equiv c_{1,1}^{(c)} - c_{1,1}^{(u)}
	= - \frac{g_1^2 \Lambda_3}{\lambda_1(\lambda_1 - \Lambda_3)}.
\end{equation}
The behavior is plotted in Fig \ref{moduliPlotChiral}.

The above result is applicable to hexagonal crystal by setting $\lambda_1 = \lambda_2$, $g_1 = g_2$ and $c_1 = c_2$ to meet the symmetry constraint.  Thalmeier \emph{et. al.}\cite{Thalmeier1991} attempted a similar analysis in the context of the hexagonal UPt$_3$, but we disagree with their result regarding $C_{66}$ (i.e. $c_{2,2}$ in our notation.)  They theoretically found a $C_{66}$ discontinuity across the upper transition, while we obtain a continuous result.  Directly below the lower transition, their result for $C_{66}$ translates to $c_{2,2} = c_2 - g_2^2/\lambda_3$ in our notation, again disagrees with our finding \eqref{chiralModuli}.  We point out that they themselves observed no $C_{66}$ discontinuity in experiment, either\cite{Thalmeier1991}.

\emph{Lower transition into nematic phase: tetragonal case.}  Now we assume that $\Lambda_2$ is the smallest, and the lower phase remain purely nematic.  Note that $\Lambda_2 \neq \Lambda_1$ is allowed only by a tetragonal crystal.  While this scenario is not immediately relevant to Sr$_2$RuO$_4$, we shall still work out the detail, because the lesson learned here is applicable to the $D_{3d}$ version of the scenario, which \emph{is} relevant to $M_x$Bi$_2$Se$_3$.

Before jumping into calculation for the lower phase, we note that $c_{2,2}^{(u)}$ from \eqref{upperModuli} vanishes at $a = a_l$ when $i = 2$ in \eqref{TcLower}.  Specifically, let $\alpha = a/a_l$, and one has
\begin{equation}
c_{2,2}^{(u)} = c_2 \frac{\Lambda_2}{\Lambda_1}(\Lambda_1-\Lambda_2)\frac{c_2}{g_2^2} \vert\alpha - 1\vert + O\!\left((\alpha-1)^2\right)
\end{equation}
directly above the upper transition.  Shear strain $\ep_2$ becomes a soft mode: we will have more comment on this shortly.

\begin{figure}
\includegraphics[width=0.48\textwidth]{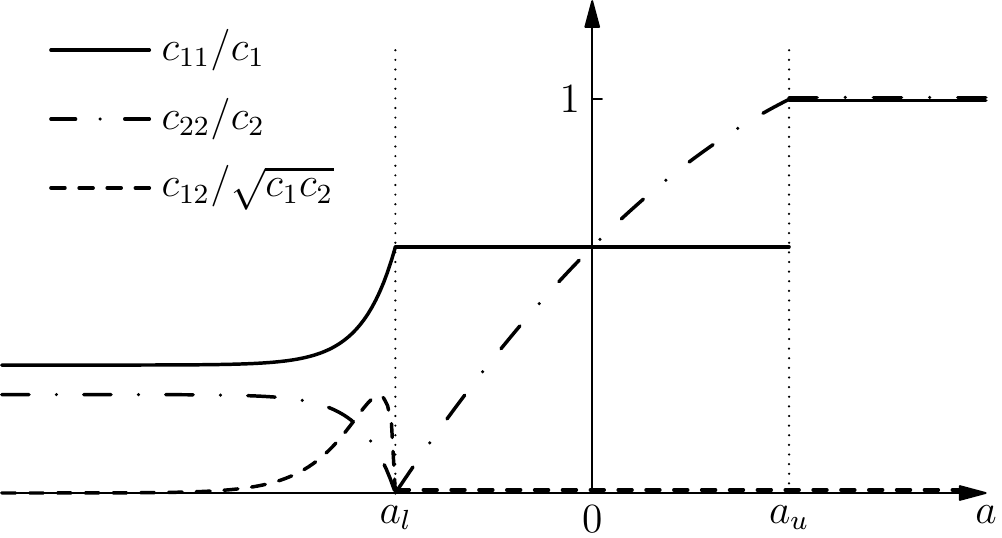}
\caption{\label{moduliPlot} Qualitative behaviors of the dimensionless $c_{i,j}/\sqrt{c_i c_j}$ across the pair of split transitions, assuming the lower phase is nematic.  The qualitative features at each critical point (kinks, discontinuity, zero, and the $\vert\delta a\vert^{1/2}$ behavior of $c_{1,2}$) are dictated by symmetry, and shared by both tetragonal and trigonal/hexagonal cases.}
\end{figure}

In principle we still follow the same procedure to calculate the elastic moduli.  The equilibrium solution is \eqref{lowerEquilibrium}.  It suffices to compute the perturbed solution to first order in $\delta \ep_i \equiv (\ep_i - \bar{\ep}_i)$ for both $i = 1, 2$.  We will refrain from printing the intermediate expressions, and will directly quote the end results:
\begin{equation}
\begin{split}
c^{(n)}_{1,1} &= c_1  - \frac{g_1^2}{\lambda_1} \left[ 1 + \frac{(\alpha^2-1)\lambda_2\Lambda_2}{(\alpha^2-1)\lambda_2(\lambda_1-\Lambda_2) +\lambda_1 g_2^2/c_2} \right]\\
c^{(n)}_{2,2} &= c_2 (\alpha^2 - 1) \,
	\frac{\Lambda_2(\lambda_1-\Lambda_2)}
		{\Lambda_2(\lambda_2-\lambda_1)+\alpha^2\lambda_2(\lambda_1-\Lambda_2)}. \\
c^{(n)}_{1,2} &= \sqrt{\alpha^2 - 1} \,
	\frac{\Lambda_2 g_1 g_2}
		{\Lambda_2(\lambda_2-\lambda_1)+\alpha^2\lambda_2(\lambda_1-\Lambda_2)},
\end{split}
\label{tetragonalNematicModuli}
\end{equation}
valid for $\alpha > 1$.  Compared with \eqref{upperModuli}, it is seen that all three components are \emph{continuous} at $\alpha = 1$ (i.e. $a = a_l$): $c_{1,1}$ shows a kink, $c_{2,2}$ \emph{vanishes} linearly with $\vert\alpha -1 \vert$ on either side of the transition, and $c_{1,2}$ grows as $\vert \alpha - 1\vert^{1/2}$ below the transition.  The general behavior of $c_{i,j}$ around $\alpha = 1$ is sketched in FIG \ref{moduliPlot}.    

This asymptotic behavior is almost obvious in hindsight: suppose one is oblivious to the superconducting aspect of the problem, the strain part of the problem resembles a structural phase transition where the horizontal two-fold rotation is spontaneously broken, and similar problems have been mapped to the transverse-field Ising model\cite{DeGennes1963, Brout1966}.  The asymptotic behavior of $c_{i,j}$ immediately follows from the well-known mean field results\cite{Stinchcombe1973}, with $\ep_2$ being a soft mode at the critical point\cite{Cochran1960}.  We expect the vanishing of $c_{2,2}$ to have a particularly dramatic impact on the ultrasonic dispersion relation.

Also immediately recognizable from the Ising model meanfield result is that the $(1,1)$-component of \emph{susceptibility} $\chi_{1,1} = (c^{-1})_{1,1} = 1/(c_{1,1}-c_{1,2}^2/c_{2,2})$ has a discontinuous step across the lower transition.  This is easily confirmed with our result \eqref{tetragonalNematicModuli}.

This is a good place to comment on the thermodynamic relation first due to Testardi\cite{Testardi1971}, relating the discontinuities in specific heat $C$ and elastic modulus $c_{i,j}$ across a second-order transition, if $\partial T_c/\partial \ep_i$ and $\partial T_c/\partial \ep_j$ are also known:
\begin{equation}
\delta c_{i,j} = -\frac{\delta C}{T_c} \frac{\partial T_c}{\partial \ep_i} \frac{\partial T_c}{\partial \ep_j}.
\label{Testardi}
\end{equation}
This relation holds approximately for both the upper transition and the lower transition into chiral phase, albeit missing the renormalization effect \eqref{renormalization}\footnote{See supplemental material for the modified formula that correctly accounts for the renormalization.  The original derivation \cite{Testardi1971} assumes a somewhat unphysical condition.}.  Given that the specific heat jump \eqref{specificHeatJumpLower}, one is tempted to invoke the \eqref{Testardi} and concludes that $c_{1,1}$ is also discontinuous.  This, of course, contradicts \eqref{tetragonalNematicModuli}.

The Testardi relation assumes the existence of a critical surface $T_c(\vec{\ep})$: all strain components must be invariant under the broken symmetry.  This is true for the two previous cases (the gauge $U(1)$ and TRS) but not true for our present case: $\ep_2 \neq 0$ in the upper phase explicitly breaks the two-fold rotational symmetry and destroys the transition.  The relation is therefore not applicable.  However, the critical line $T_c(\ep_1)$ is well-defined if one allows $\ep_2$ to reach equilibrium: $\ep_2 = \bar{\ep}_2(\ep_1)$.  The Testardi relation would instead give the discontinuity in $1/\chi_{1,1}$, though it still misses the renormalization \eqref{renormalization}.

One interesting limit is $g_2 \rightarrow 0$.  Since $\ep_2$ is decoupled from the problem, $c_{1,1} \rightarrow 1/\chi_{1,1}$ and become discontinuous at the transition.  This is indeed readily seen in \eqref{tetragonalNematicModuli}.  Mathematically, $\lambda_1 g_2^2/c_2$ plays the some role as $\Lambda_1 \Delta$ did for $c_{2,2}^{(u)}$ in \eqref{upperModuli}, smoothing out the singularity.  Since the renormalization $g_2^2/c_2$ is likely to be small compared to $\lambda_2$, in practice $c_{1,1}$ would still exhibit a ``near-step'':
\begin{equation}
\delta_{n} c_{1,1} \approx - \frac{g_1^2}{\lambda_1}\frac{\Lambda_2}{\lambda_1 - \Lambda_2},
\end{equation}
and the width of this broadened step is of the order of
\begin{equation}
\Delta\alpha = \frac{1}{2}\frac{g_2^2}{c_2 \lambda_2}  \frac{\lambda_1}{\lambda_1-\Lambda_2}.
\end{equation}

\emph{Trigonal nematic superconductor.}  The $M_x$Bi$_2$Se$_3$ crystal has $D_{3d}$ symmetry that forces $\Lambda_1 = \Lambda_2 \equiv \Lambda$, $c_1 = c_2 \equiv c$ and $g_1 = g_2 \equiv g$.  Without the SBF $\Delta$, the free energy \eqref{GLfull} becomes invariant under arbitrary rotation: the theory enjoys an emergent $O(2)$ symmetry.  It is assumed that the material prefers the nematic state and consequently $\Lambda < \Lambda_3$.  The possible $O(\vert\eta\vert^6)$ term, however, explicitly breaks this $O(2)$ symmetry down to six-fold:
\begin{equation}
\fe_6 = \Gamma_1 \vert\eta\vert^6
+ \Gamma_2 \left(\langle\sigma\rangle_1^3 - 3 \langle\sigma\rangle_2^2 \langle\sigma\rangle_1\right).
\label{GL6th}
\end{equation}
The coefficient $\Gamma_2 < 0$ favors $\vec{\eta} \propto (1, 0)$ and the other two equivalent directions, while $\Gamma_2 > 0$ favors $\vec{\eta} \propto (0, 1)$ and equivalent directions.  The sign of $\Gamma_2$ is undetermined.

The original GL theory \eqref{GLoriginal} is self-consistent in that it completely accounts for the leading $O(a^2)$ behavior of the free energy.  We demand that $\fe_6 \sim O(a^3)$ remains a small perturbation, so the modified GL theory remains applicable.  Specifically, let $D_c = \vert \bar{\eta}\vert^2$ at lower transition, we demand $\Gamma_i D_c \ll \lambda$ for $i = 1,2$.  Since the possible lower transition is caused by competition between $\fe_6$ and $\Delta$, it may remain within the range of validity of our theory yet.

The SBF $\Delta$ breaks the crystal symmetry down to $C_{2h}$, but the important point is one horizontal two-fold rotational axis remains.  The sign of $\Delta$ determines whether $\vec{\eta}$ in the upper phase is along $(1,0)$ or $(0,1)$.  If it doesn't match the preference dictated by $\Gamma_2$, a lower transition qualitatively similar to the tetragonal case takes place.  The phase diagram of this model has been thoroughly discussed by the present authors\cite{How2019}.  Experimentally, the sign of $\Delta$ has been reported to be sample-dependent\cite{Du2017, Kuntsevich2019, Kawai2020}: \emph{some} samples must exhibit this lower transition automatically.  It what follows, we assume both $\Gamma_2$ and $\Delta$ to be \emph{positive}; the case of negative $\Gamma_2$ can be obtained by some trivial sign flips.

The upper critical temperature is again $a_u = \Delta$.  The upper phase equilibrium solution and moduli for trigonal crystal are found:
\begin{equation}
\begin{split}
\vert\bar{\eta}\vert^2 &= 
	\frac{-\Lambda + \sqrt{\Lambda^2 - 12(\Gamma_1 + \Gamma_2)(a-\Delta)}}{6(\Gamma_1 + \Gamma_2)}, \\
c^{(ut)}_{1,1} &= c - \frac{g^2}{\lambda + 6(\Gamma_1 + \Gamma_2)\vert\bar{\eta}\vert^2}, \\
c^{(ut)}_{2,2} &= c - \frac{g^2 \vert\bar{\eta}\vert^2}{\Delta + (g^2/c)\vert\bar{\eta}\vert^2 - 9\Gamma_2 \vert\bar{\eta}\vert^4},
\end{split}
\label{trigonalUpperModuli}
\end{equation}
and $c^{(ut)}_{1,2} = \bar{\theta} = \bar{\chi} = 0$.  One sees that \eqref{trigonalUpperModuli} and \eqref{upperModuli} coincides when $a \rightarrow \Delta^{-}$: the additional $\fe_6$ is of no importance when $\vert\eta\vert$ is vanishingly small.  In particular, the discontinuity in $c_{1,1}$ is still given by \eqref{upperStep}.  We also note that the renormalization \eqref{renormalization} must be kept in order to obtain a sensible answer for $c_{2,2}$ in the $\Delta \rightarrow 0$ limit: without the renormalization $c_{2,2}$ goes to negative infinity below the transition (utterly unphysical), but in \eqref{trigonalUpperModuli} it merely becomes soft.  As advertised above, the distinction between $\lambda$ and $\Lambda$ cannot be ignored for the trigonal problem.

Unfortunately, analytic solution cannot be obtained for the lower phase, and we resort to expanding in small $\delta t \equiv (a-a_l)/a_l$.  Full expressions for leading order term can be obtained\footnote{See supplemental material for the full expressions}.  But we will further invoke the condition $\Gamma_i D_c \ll \lambda$, and also assume that the renormalized $\Lambda$ is of the same order as the bare $\lambda$.  One then goes through similar calculation and obtains in the lower nematic phase for trigonal crystal:
\begin{equation}
\begin{split}
D_c &= \frac{1}{3} \sqrt{\frac{\Delta}{\Gamma_2}}, \\
a_l &= -\frac{1}{3}\left[\Lambda \sqrt{\frac{\Delta}{\Gamma_2}} +
\left(\Gamma_1 - 2\Gamma_2\right)\frac{\Delta}{\Gamma_2}\right]
	\approx -\Lambda D_c, \\
c_{1,1}^{(nt)} &\approx c\left[ 1 - \left(\frac{g^2}{c\Gamma_2 D_c}\right) \! \left(\frac{\Lambda}{\lambda}\right) \! \left(\frac{g^2/(c\Lambda) + 3\delta t/2}{g^2/(c\Gamma_2 D_c) + 36 \delta t} \right) \right], \\
c_{2,2}^{(nt)} &\approx c\left[ 1 - \left(\frac{g^2}{c\Gamma_2 D_c}\right) \! \left(\frac{\Lambda}{\lambda}\right) \! \left(\frac{\lambda/\Lambda + 3\delta t/2}{g^2/(c\Gamma_2 D_c) + 36 \delta t} \right) \right], \\
c_{1,2}^{(nt)} &\approx \left(\frac{g^2}{c\Gamma_2 D_c}\right) \! \left(\frac{\Lambda}{\lambda}\right) \! \left(\frac{\sqrt{3\delta t/2}}{g^2/(c\Gamma_2 D_c) + 36 \delta t} \right).
\end{split}
\label{trigonalNematicModuli}
\end{equation}
The quantity $g^2/(c\Gamma_2 D_c)$ is the ratio between two characteristic scales: $g^2/c$ for the strain renormalization, and $\Gamma_2 D_c$ characterizing the lower transition.  We have required $\Gamma_2 D_c \ll \lambda$, but typically the renormalization is also small: $g^2/c \ll \lambda$.  Their ratio is not necessarily big or small, and is not constrained by the validity of the GL theory.  Thus we are justified in retaining the ratio in \eqref{trigonalNematicModuli}.  However, if the renormalization effect is summarily ignored, both $c_{2,2}^{(nt)}$ and $c_{1,2}^{(nt)}$ would diverge at the lower critical point: again  unphysical.  We stress again that the renormalization \eqref{renormalization} must be kept.

Although the detailed expressions become way more complicated, the qualitative physics of this lower transition is still governed by the spontaneous breaking of the two-fold horizontal rotational symmetry: $c_{1,1}$ exhibits a kink, $c_{1,2}$ grows as $\sqrt{\delta t}$ in the lower phase, and $c_{2,2}$ \emph{vanishes} at the transition, as sketched in Fig \ref{moduliPlot}.  The (inverse) susceptibility $1/\chi_{1,1}$ is discontinuous, and the step height can be related to the specific heat discontinuity using the Testardi thermodynamic\eqref{Testardi}, up to necessary corrections\cite{Note2}, but now there is no clear limit where it can be equated with $c_{1,1}$.

Our result has important implication to the supposed nematic superconductivity in $M_x$Bi$_2$Se$_3$.  The ``nematic hypothesis'' was first proposed as the origin for the two-fold anisotropy responses that are incompatible with the lattice symmetry and observed only in the superconducting phase\cite{Fu2014, Matano2016, Pan2016}.  The persistence of the two-fold direction, however, requires a pinning SBF that explicitly breaks the lattice symmetry in the normal state\cite{Kuntsevich2018, How2019}.  If one accepts the necessity of an explicit SBF, \emph{the nematic superconductivity is no longer mandated by symmetry of the problem.}  In addition, ARPES result\cite{Lahoud2013} confirms that the conduction band dispersion remains very similar to the undoped crystal\cite{Zhang2009, Liu2010}, and would indicate that the normal state behaves as a weakly-interacting Fermi liquid.  One is left with a dilemma: either there is some exotic phonon pairing mechanism that explicitly disfavors the s-wave pairing\cite{Brydon2014, Wan2014, Wu2017, Wang2019} (c.f. \cite{Nomoto2020}), or the superconductivity is single-component (and therefore \emph{not nematic}) yet unusually susceptible to directional perturbations.  To our knowledge, no existing experiments is able to differentiate the two possibilities, but the two scenarios suggest vastly different microscopic models.

Two key signatures for the nematic scenario are proposed here.  First, the discontinuity in $c_{11}$ across the upper transition is unique to a multi-component order parameter.  Moreover, the \emph{existence} of a lower transition constitutes a proof beyond all doubt.  The present authors suggested that the lower transition may be identified with calorimetric measurements\cite{How2019}, but we concede here that the specific heat discontinuity is proportional to $\sqrt{\Delta}$, and may well be too small to be observed.  On the other hand, the \emph{vanishing} of $c_{2,2}$ at the lower transition is a very dramatic signal, one that we hope would be easily picked up by, says, ultrasound experiments.  As discussed above, the sign of the SBF $\Delta$ varies sample-to-sample: some samples must exhibit a lower transition \emph{without external manipulation.}

\emph{Conclusion.}  In this paper we examine the shear modulus anomaly of an unconventional superconductor in the presence of an SBF.  The hallmark of such systems is a pair of split superconducting transitions, and we work out the anomaly for both the upper and lower transitions within the GL theory.

Indeed, for both Sr$_2$RuO$_4$ and $M_x$Bi$_2$Se$_3$, two materials that are proposed to host unconventional superconductivity, the experimental evidence of split transitions has been inconclusive, to say the least.  In light of recent ultrasound experiments, we hope that the measurement of shear moduli will prove fruitful in the experimental search of the lower transitions.  This is especially the case for $M_x$Bi$_2$Se$_3$, where we predict that one of the shear modes must become soft, i.e. its modulus vanishes, at the lower transition.  We believe such a dramatic signature cannot go unnoticed.

\begin{acknowledgments}
This work was supported by Academia Sinica through AS-iMATE-109-13, and the Ministry of Science and Technology (MOST), Taiwan through MOST-107-2112-M-001-035-MY3.
\end{acknowledgments}

\bibliography{shearModulus}

\onecolumngrid

\section*{Supplemental Material}

\subsection*{Calculation of shear modulus}

The shear modulus $c_{ij}$ is in principle defined by $c_{ij} = \partial^2 F/ \partial\ep_i \partial \ep_j$ as given in the main text.  In practice, though, there are many different ways to obtain the desired result.  We describe our own method here.

Define the stress
\begin{equation}
\tau_i(\vec{\ep}; \vec{\eta}) \equiv \frac{\partial \fe}{\partial \ep_i};
\end{equation}
it is the ``restoring restoring'' the system felt when $\ep_i$ is taken out of equilibrium.  Also define the $\tilde{\eta}(\vec{\ep})$ to solve the GL equation $\partial \fe/\partial\vec{\eta} = 0$ in the presence of arbitrary $\vec{\ep}$.  Then the modulus will be given by
\begin{equation}
c_{ij} = \frac{d}{d\ep_j} \tau_i(\vec{\ep}, \tilde{\eta}(\vec{\ep})),
\end{equation}
the derivative to be evaluated at equilibrium.

Let us parameterize $\vec{\eta} = \sqrt{D}(\cos \theta, e^{i\chi} \sin \theta)$.  Given the strain part $\fe_{\ep}$, we get
\begin{equation}
\begin{split}
\tau_1 &= -g_1 D \cos 2\theta + c_1 \ep_1, \\
\tau_2 &= -g_2 D \sin 2\theta \cos \chi + c_2 \ep_2,
\end{split}
\end{equation}
and the modulus becomes
\begin{equation}
\begin{pmatrix}
c_{11} & c_{12} \\
c_{21} & c_{22}
\end{pmatrix}
= 
\begin{pmatrix}
c_1 & 0 \\
0 & c_2
\end{pmatrix}
-
\begin{pmatrix}
g_1 \cos 2\theta & -g_1 \sin 2\theta \\
g_2 \sin 2\theta \cos \chi & g_2 \cos 2\theta \cos \chi
\end{pmatrix}
\begin{pmatrix}
\frac{\partial D}{\partial \ep_1} & \frac{\partial D}{\partial\ep_2} \\
D \frac{\partial 2\theta}{\partial \ep_1} & D \frac{\partial 2\theta}{\partial\ep_2} \\
\end{pmatrix}
+ g_2 D \sin 2\theta \sin \chi \begin{pmatrix}
0 &  0 \\
\frac{\partial \chi}{\partial \ep_1} & \frac{\partial \chi}{\partial \ep_2}.
\end{pmatrix}
\label{SK}
\end{equation}
The dependence of $D$, $\theta$, $\chi$ on $\vec{\ep}$ is determined from the Ginzburg-Landau equations $\partial \fe/\partial D = \partial \fe /\partial \theta = \partial \fe / \partial \chi = 0$.  But it turns out that the inverse of the matrix is a lot easier to compute:
\begin{equation}
\begin{pmatrix}
\frac{\partial \ep_1}{\partial D} & \frac{1}{D} \frac{\partial \ep_1}{\partial 2\theta}\\
\frac{\partial\ep_2}{\partial D} & \frac{1}{D} \frac{\partial\ep_2}{\partial 2\theta} \\
\end{pmatrix}
=
\begin{pmatrix}
\frac{\partial D}{\partial \ep_1} & \frac{\partial D}{\partial\ep_2} \\
D \frac{\partial 2\theta}{\partial \ep_1} & D \frac{\partial 2\theta}{\partial\ep_2} \\
\end{pmatrix}^{-1}
\label{derivativeMatrix}
\end{equation}

As an illustration, let us consider the lower nematic phase for the trigonal crystal, i.e. we assume $\Lambda < \Lambda_2$, and look for the phase where at equilibrium $\theta \neq 0$.  It can be easily shown that $\chi = \partial \chi/\partial \ep_i = 0$ for this case, and one only needs to treat $D$ and $\theta$.  In the presence of $\vec{\ep}$, the GL equations for $D$ and $\theta$ yields:
\begin{equation}
\begin{split}
g_1 \ep_1 &= - \Delta + a \cos 2\theta + \lambda D \cos 2\theta + 3 \Gamma_1 D^2 \cos 2\theta + 3 \Gamma_2 D^2 \cos 4\theta, \\
g_2 \ep_2 &= a \sin 2\theta + \lambda D \sin 2\theta + 3 \Gamma_1 D^2 \sin 2\theta - 3 \Gamma_2 D^2 \sin 4\theta.
\end{split}
\end{equation}
As advertised, the matrix \eqref{derivativeMatrix} is easily calculated from here, and its inverse is taken trivially.

After the derivatives are taken, one must insert the equilibrium solutions $\bar{D}$ and $\bar{\theta}$.  These comes from solving the GL equations with the condition $\partial \fe/\partial \vec{\ep} = 0$.  An analytic solution is not possible, and we resort to expansion in $\delta t = (a-a_l)/a_l$:
\begin{equation}
\begin{split}
\bar{D} &= D_c + \frac{8}{3} D_c\, \bar{\theta}^2 + O(\bar{\theta}^4), \\
\bar{\theta}^2 &= \frac{-a_l \, \delta t}{\frac{8}{3} \Lambda D_c +16 \Gamma_1 D_c^2 - 20 \Gamma_2 D_c^2}  + O(\delta t^2).
\end{split}
\end{equation}
We now have all ingredients necessary to evaluate the moduli using \eqref{SK}.  The result is
\begin{equation}
\begin{split}
c_{11} &= c \left\lbrace 1 - \frac{g^2/c - \bar{\theta}^2 (4\Lambda + 24 \Gamma_1 D_c - \Gamma_2 D_c)}{L}\right\rbrace, \\
c_{22} &= c \left\lbrace 1 - \frac{[\lambda + 6(\Gamma_1+\Gamma_2)D_c] - \bar{\theta}^2 (4\Lambda - 8\Gamma_1 D_c - 28 \Gamma_2 D_c)}{L}\right\rbrace, \\
c_{12} &= \frac{\bar{\theta} (\Lambda + 12 \Gamma_1 D_c - 24\Gamma_1 D_c)}{L},
\end{split}
\end{equation}
where the common denominator $L$ is
\begin{equation}
L = \lambda + 6(\Gamma_1 + \Gamma_2)D_c + \bar{\theta}^2
\left[4(4\Gamma_1-13\Gamma_2)D_c + 48 (c \Gamma_2 D_c/g^2) (2\lambda + 12\Gamma_1 D_c-15\Gamma_2 D_c)\right].
\end{equation}

\subsection*{Thermodynamic relation done right}

A thermodynamic relation between discontinuities in specific heat and elastic moduli at a second order phase transition, first due to Testardi[44], is often quoted in the literature.  We want to show that the Testardi relation does not exactly apply to typical experimental conditions, and how it can be modified to be more applicable.

Consider a system that lives in an $(N+1)$-dimensional space, with temperature $T$ and $N$ strain components $\ep_1, \dots \ep_N$, collectively denoted $\vec{\ep}$.  There are two phases (\emph{upper} and \emph{lower}), separated by a surface of second order phase transition which is given by $T = T_c(\vec{\ep})$.

The Testardi relation states the following.  Let the specific heat discontinuity across the critical surface be given by
\begin{equation}
\frac{\Delta C_V}{T_c(\vec{\ep})} = -\alpha(\vec{\ep}).
\label{specificHeatJumpTestardi}
\end{equation}
Then the discontinuity of elastic moduli $c_{i, j}$ will be
\begin{equation}
\Delta c_{i,j} = \alpha(\vec{\ep})\, \frac{\partial^2 T_c}{\partial\ep_i\partial\ep_j}.
\label{modulusJumpTestardi}
\end{equation}
Here the difference is defined to be the upper phase value minus the lower phase value.

We proceed to carefully review the derivation of this result.  Let $\fe^U(T, \vec{\ep})$ and $\fe^L(T,\vec{\ep})$ denotes the free energy of the upper and lower phases, respectively.  The assumption that the phase transition is of second order implies that $\fe$ and its first temperature derivative must be continuous across the transition.  One therefore expects the asymptotic form
\begin{equation}
\fe^U(T, \vec{\ep}) - \fe^L(T, \vec{\ep}) \approx \frac{1}{2}\,\alpha(\vec{\ep}) \, (T - T_c(\vec{\ep}))^2.
\label{fDifference}
\end{equation}
to hold immediately below the critical surface.  By taking second derivatives of \eqref{fDifference}, the relations \eqref{specificHeatJumpTestardi} and \eqref{modulusJumpTestardi} immediately follow.

The heat capacity in question here, $C_V = - T\,(\partial^2 \fe/\partial T^2)\vert_{\vec{\ep}}$, is measured under the constraint of no deformation.  When a phase transition has a structural nature, this is seldom the actual experimental condition.  Instead, the strains are usually not specifically controlled in a calorimetric experiment and largely left to equilibrate by themselves.  The heat capacity so measured would correspond to $C = - T\,(d/dT)^2 \fe(T, \bar{\ep}(T))$, where $\bar{\ep}(T)$ describes the equilibrium state at given $T$.

To illustrate the point, consider the very simple example of a scalar order parameter $\psi$ coupled to a scalar strain $\ep$.  The GL free energy reads:
\begin{equation}
\fe = (a - g \ep)\vert\psi\vert^2 + \frac{1}{2} b \vert\psi\vert^4 + \frac{1}{2} c_0 \ep^2.
\end{equation}
It is easily identified that $T_c(\ep) = T_c(0) + g\ep$.  Standard calculations give $\Delta C = T_c(0)/(b - g^2/c_0)$ and $\Delta c = -g^2/b$.  Testardi relation \eqref{specificHeatJumpTestardi} and \eqref{modulusJumpTestardi} fail to account for the renormalization of $b$ even in this simple example.

Let us go back to the $N$-component strain case, and see if the Testardi relation can be modified to reflect the experimental condition. The physical trajectory of the system in the upper phase is given by $\vec{\ep} = \bar{\ep}^U(T)$ that solves $\partial \fe^U/\partial \vec{\ep} = 0$.  At some $T = T_0$ the path intersect with the critical surface.  The curve $\bar{\ep}^U(T)$ smoothly continues for $T < T_0$, representing an unstable equilibrium state.  The stable equilibrium below the critical surface follows the curve $\vec{\ep} = \bar{\ep}^L(T)$, solution of $\partial \fe^L/\partial \vec{\ep} = 0$.  Let $\delta\vec{\ep} = \bar{\ep}^L(T) - \bar{\ep}^U(T)$, and consider the quantity:
\begin{equation}
\Delta \fe = \fe^U(T, \bar{\ep}^U) - \fe^L(T, \bar{\ep}^L) \approx \frac{1}{2}\,\alpha(\bar{\ep}^L) \, (T - T_c(\bar{\ep}^L)))^2 - \frac{1}{2}\sum_{i,j} c^U_{i,j} \delta\ep_i \delta\ep_j.
\label{fDifference2}
\end{equation}
One first expands $\fe^U(T, \bar{\ep}^L) = \fe^U(T, \bar{\ep}^U + \delta\vec{\ep})$ in $\delta\vec{\ep}$, and then makes use of \eqref{fDifference2} to obtain the RHS.  By construction, the coefficient $c^U_{i,j}$ is the elastic modulus of the upper phase.  The experimentally measured heat capacity step is in fact:
\begin{equation}
\frac{\Delta C}{T_0} = -\frac{d^2 \Delta \fe}{dT^2} 
= - \alpha \left[ 1 + \left(\sum_i \frac{\partial T_c}{\partial\ep_i} \frac{\partial \bar{\ep}^L_i}{\partial T} \right)^2\right] - c^U_{i,j} \frac{d \delta \ep_i}{d T} \frac{d \delta \ep_j}{d T},
\end{equation}
all derivatives evaluated at the critical point.  The correction terms that were omitted in \eqref{specificHeatJumpTestardi} aqccounts for the discrepancy between bare and renormalized quartic coupling.

Finally we note that the thermodynamic relation rests on the existence of $T_c(\vec{\ep})$.  The strain $\vec{\ep}$ is required to transform trivially under the broken symmetry, else there will not even be a phase transition at non-zero $\vec{\ep}$.  Thus the relation is simply not applicable to the lower transition to nematic state discussed in the main text.

\end{document}